\documentclass[superscriptaddress,onecolumn,secnumarabic,
amssymb,amsmath,nobibnotes,aps,prd,showkeys,showpacs,nofootinbib]{revtex4}

\usepackage[latin1]{inputenc}
\usepackage{graphicx}
\usepackage[english]{babel}

\usepackage{amsmath}
\usepackage{amssymb}
\usepackage{amsfonts}
\usepackage{colordvi}
\usepackage{psfrag}
\usepackage{color}

\begin{document}
\def\cL{{\cal L}}
\def\be{\begin{equation}}
\def\ee{\end{equation}}
\def\bea{\begin{eqnarray}}
\def\eea{\end{eqnarray}}
\def\beq{\begin{eqnarray}}
\def\eeq{\end{eqnarray}}
\def\tr{{\rm tr}\, }
\def\nn{\nonumber \\}
\def\e{{\rm e}}
\def\bef{\begin{figure}}
\def\eef{\end{figure}}
\newcommand{\ans}{ansatz }
\newcommand{\eeqn}{\end{eqnarray}}
\newcommand{\bd}{\begin{displaymath}}
\newcommand{\ed}{\end{displaymath}}
\newcommand{\mat}[4]{\left(\begin{array}{cc}{#1}&{#2}\\{#3}&{#4}
\end{array}\right)}
\newcommand{\matr}[9]{\left(\begin{array}{ccc}{#1}&{#2}&{#3}\\
{#4}&{#5}&{#6}\\{#7}&{#8}&{#9}\end{array}\right)}
\newcommand{\matrr}[6]{\left(\begin{array}{cc}{#1}&{#2}\\
{#3}&{#4}\\{#5}&{#6}\end{array}\right)}
\newcommand{\cvb}[3]{#1^{#2}_{#3}}
\def\lsim{\raise0.3ex\hbox{$\;<$\kern-0.75em\raise-1.1ex
e\hbox{$\sim\;$}}}
\def\gsim{\raise0.3ex\hbox{$\;>$\kern-0.75em\raise-1.1ex
\hbox{$\sim\;$}}}
\def\abs#1{\left| #1\right|}
\def\simlt{\mathrel{\lower2.5pt\vbox{\lineskip=0pt\baselineskip=0pt
           \hbox{$<$}\hbox{$\sim$}}}}
\def\simgt{\mathrel{\lower2.5pt\vbox{\lineskip=0pt\baselineskip=0pt
           \hbox{$>$}\hbox{$\sim$}}}}
\def\unity{{\hbox{1\kern-.8mm l}}}
\newcommand{\eps}{\varepsilon}
\def\ep{\epsilon}
\def\ga{\gamma}
\def\Ga{\Gamma}
\def\om{\omega}
\def\omp{{\omega^\prime}}
\def\Om{\Omega}
\def\la{\lambda}
\def\La{\Lambda}
\def\al{\alpha}
\newcommand{\ov}{\overline}
\renewcommand{\to}{\rightarrow}
\renewcommand{\vec}[1]{\mathbf{#1}}
\newcommand{\vect}[1]{\mbox{\boldmath$#1$}}
\def\tm{{\widetilde{m}}}
\def\mcirc{{\stackrel{o}{m}}}
\newcommand{\Dm}{\Delta m}
\newcommand{\dm}{\varepsilon}
\newcommand{\tanb}{\tan\beta}
\newcommand{\nbar}{\tilde{n}}
\newcommand\PM[1]{\begin{pmatrix}#1\end{pmatrix}}
\newcommand{\up}{\uparrow}
\newcommand{\down}{\downarrow}
\def\omE{\omega_{\rm Ter}}
%

\newcommand{\Dsusy}{{susy \hspace{-9.4pt} \slash}\;}
\newcommand{\DCP}{{CP \hspace{-7.4pt} \slash}\;}
\newcommand{\mc}{\mathcal}
\newcommand{\gr}{\mathbf}
\renewcommand{\to}{\rightarrow}
\newcommand{\gtc}{\mathfrak}
\newcommand{\wh}{\widehat}
\newcommand{\br}{\langle}
\newcommand{\kt}{\rangle}


\def\lsim{\mathrel{\mathop  {\hbox{\lower0.5ex\hbox{$\sim$}
\kern-0.8em\lower-0.7ex\hbox{$<$}}}}}
\def\gsim{\mathrel{\mathop  {\hbox{\lower0.5ex\hbox{$\sim$}
\kern-0.8em\lower-0.7ex\hbox{$>$}}}}}

\def\nn{\\  \nonumber}
\def\de{\partial}
\def\brf{{\mathbf f}}
\def\bbf{\bar{\bf f}}
\def\bF{{\bf F}}
\def\bbF{\bar{\bf F}}
\def\bA{{\mathbf A}}
\def\bB{{\mathbf B}}
\def\bG{{\mathbf G}}
\def\bI{{\mathbf I}}
\def\bM{{\mathbf M}}
\def\bY{{\mathbf Y}}
\def\bX{{\mathbf X}}
\def\bS{{\mathbf S}}
\def\bb{{\mathbf b}}
\def\bh{{\mathbf h}}
\def\bg{{\mathbf g}}
\def\bla{{\mathbf \la}}
\def\bmu{\mathbf m }
\def\by{{\mathbf y}}
\def\bmu{\mbox{\boldmath $\mu$} }
\def\bsig{\mbox{\boldmath $\sigma$} }
\def\bunity{{\mathbf 1}}
\def\cA{{\cal A}}
\def\cB{{\cal B}}
\def\cC{{\cal C}}
\def\cD{{\cal D}}
\def\cF{{\cal F}}
\def\cG{{\cal G}}
\def\cH{{\cal H}}
\def\cI{{\cal I}}
\def\cL{{\cal L}}
\def\cN{{\cal N}}
\def\cM{{\cal M}}
\def\cO{{\cal O}}
\def\cR{{\cal R}}
\def\cS{{\cal S}}
\def\cT{{\cal T}}
\def\eV{{\rm eV}}

\title{Topological portals from matter to antimatter}

\author{Andrea Addazi}

\affiliation{ Center for Theoretical Physics, College of Physics Science and Technology, Sichuan University, 610065 Chengdu, China}
\affiliation{Laboratori Nazionali di Frascati INFN Via Enrico Fermi 54, I-00044 Frascati, Italy}

\date{\today}

\begin{abstract}
We discuss possibilities of generating 
a Majorana mass for the neutron from topological quantum gravity effects which survive at mesoscopic scales from decoherence.
We show how virtual micro-black hole (BH) pairs with skyrme/baryon hairs
induce a neutron-antineutron transition which can be tested in next generation of experiments. 
Such effects do not destabilize the proton.
We also discuss how BHs with mix ordinary and mirror baryon hairs can mediate 
neutron-mirror neutron mixings. 
\end{abstract}
\keywords{Neutron-antineutron physics, mirror matter, dark matter, Quantum gravity, instantons, black hole hairs.}

\maketitle

\section{Introduction and main ideas}

Kuzmin's idea of neutron-antineutron physics \cite{Kuzmin} is more actual and exciting than ever in light of next experimental possibilities \cite{Addazi:2020nlz}.
Effective field theory would suggest to relate a
neutron-antineutron transition, generated by a 
neutron Majorana mass $\epsilon\,nn+h.c$, to $D=9$ six quarks $|\Delta B|=2$ violating effective operators
$(udd)^{2}/\Lambda^{5}$ or $(qqd)^{2}/\Lambda^{5}$. 
Here, $\Lambda$ is the new physics scale and, as it is well known, 
it is related to the effective Majorana mass as 
$\epsilon\simeq \Lambda_{QCD}^{6}/\Lambda^{5}$.
The current best limits on $n-\bar{n}$ set an exclusion scale of $\Lambda>300\, {\rm TeV}$ \cite{BC}
and there are future possibilities to push it towards $1000\, {\rm TeV}$ or so
\cite{Addazi:2020nlz,Phillips:2014fgb}. 
Thus, for generating a neutron Majorana mass testable in next future, we would need for new heavy fields 
having a combination of their masses on a multi-hundred TeV scale. 
For example, there are models with heavy vector-like colored scalar triplets
and a sterile neutron see-saw partner \cite{Berezhiani:2005hv,Berezhiani:2009ldq,Addazi:2015ata,Berezhiani:2015afa,Addazi:2016rgo} or 
GUT inspired scenarios involving colored scalar sextets \cite{sextets,Mohapatra:1986dg,Babu:2013yca,Babu:2006xc}. 
Attempts to connect UV completion of the baryon violating six-quark operators 
with post-sphaleron baryogenesis were also proposed \cite{Mohapatra:1986dg,Babu:2013yca,Babu:2006xc}. 
An exciting possibility is to consider a $n-\bar{n}$ oscillation 
from neutron-mirror neutron $(n-n')$ transitions ($|\Delta B=1|$)  \cite{Berezhiani:2005hv,Berezhiani:2009ldq,Mohapatra:2005ng,Berezhiani:2011da,Berezhiani:2017azg,Berezhiani:2020vbe,Babu:2021mjg}. 
Also in this case, $n-n'$ would be generated by
effective $D=9$ operators $(udd)(u'd'd')/\Lambda'^{5}$ or $(qqd)(q'q'd')/\Lambda'^{5}$
with a new physics scale considerably lower than the $n-\bar{n}$ one \cite{Berezhiani:2005hv,Berezhiani:2009ldq,Mohapatra:2005ng,Berezhiani:2011da,Berezhiani:2015afa,Berezhiani:2017azg,Berezhiani:2020vbe,Babu:2021mjg}.

Here, we will move on an alternatively radical path: we will assume 
that no any new heavy particle states are necessary for a neutron-antineutron transition. We propose that $n-\bar{n}$ can be obtained in Standard Model (SM)
from non-perturbative quantum gravity effects
surviving at {\it mesoscopic length scales} much larger than the Planck length $L_{Pl}$.  
It is a commonly accepted argument that quantum gravity effects
mediate transitions violating any global symmetry such as Baryon/Lepton number conservations
\cite{G1,G2,G3,G4}. 
Nevertheless, these are expected to disappear at energies much below the Planck mass $M_{Pl}$. 
 Apparently, in $L_{Pl}\rightarrow 0$ limit, there is no any way out from the fact that (minimal) SM
cannot generate any $B-L$ violations, at both perturbative and non-perturbative level \cite{VW}.
Indeed, while the perturbative lagrangian of the SM explicitly preserves $B$ and $L$ 
as accidental number conservations, electroweak non-perturbative effects such as sphalerons 
can only violate $B+L$, but not $B-L$, around the electroweak phase transition in the early Universe \cite{Sphalerons}. 
Nevertheless, sphalerons are exponentially suppressed in late Universe and practically untestable in low energy physics \cite{Sphalerons}. 
From strong QCD effects we would also not expect any surprising baryon violations, 
if the Vafa-Witten theorem \cite{VW} was not violated by any subtle dynamical effects. An interesting conjecture proposed by
{\it Berezhiani} is that an effective supersymmetrization of d.o.f in the Fermi world may contradict the VW-theorem \cite{Berezhiani:2015afa}.
On the other hand, quantum gravity effects,
although not well known around the Planck scale,
are more theoretically controllable in low-energy regimes. 
Nevertheless, subtle and unexpected residual memory effects may survive at larger scales
from quantum dechoerence attacks if there was any {\it quantum amplification mechanism}.

Indeed, having defined $\bar{L}$ as the length scale in renormalization group approach, there are several quantum gravity processes that do not scale as powers of 
$(L_{Pl}/\bar{L})^2$ and they can be interpreted as coherent semi-classical states. 
This is the case of gravitational instantons, 
that are classical saddle solutions
of the euclidean path integral \cite{I1,I2,I3,I4}. 
As it is know, instantons are generically interpreted as semi-classical and non-perturbative 
quantum tunneling processes in real space-time. 
Indeed, there is a correspondence among instantons and solitons (see for Example Ref.\cite{Addazi:2016yre}).
For example, in non-abelian Yang-Mills theories coupled to a scalar Higgs field, instantons can be interpreted 
as tunnelings of monopoles or antimonopoles out from domain walls or a Josephson junction \cite{Dvali:2007nm}.
In case of gravity, gravitational instanton dominating the euclidean path integral corresponds to  
a tunneling process of a Black Hole (BH) into a White Hole (WH), i.e. a virtual Wormhole \cite{I1,I2,I3,I4}. 
As a quantum fluctuation of space-time, a gravitational instanton describes the appearence of a virtual BH pair. 
As every instantons, also gravitational instantons correspond to transition probabilities
with a WKB exponential suppression as $e^{-I_{E}}$, where $I_{E}$ is the Euclidean action 
of the instantons. 
 Nevertheless, there is an important difference in gravitational case: 
 the euclidean action is proportional to the BH entropy in correspondence to the 
 instanton solution. 
 Indeed, this is related to the fact that the euclidean quantum gravity path integral,
 in semiclassical regime, corresponds  to the thermodynamical partition function of 
BH \cite{I1,I2,I3,I4,I5}. Thus,
 all thermodynamical quantities of the BH can be defined, including Bekeinstein-Hawking 
 temperature, entropy, free-energy and so on. Now, as it is notoriously known, 
 BH entropy scales as the BH Area rather than Volume
 and thus also the euclidean action of gravitational instantons is holographic:
 \begin{equation}
 \label{SEE}
S\sim I_{E}\sim (R_{BH}/L_{Pl})^{2}+O({\rm log}R_{BH}/L_{Pl})\sim N+O({\rm log}N)\, ,
 \end{equation}
  where $I_E$ is the Euclidean gravitational action, $S$ is the BH entropy, $R_{BH}$ is the BH radius, $N$ is the number of {\it quantum hairs} or {\it planckian qubit} stored on the BH surface 
  \cite{QH1,QH1b,QH2,DG} while from the instanton side it may correspond to the topological winding number \cite{Addazi:2020mnm}
  (see also Ref.\cite{Addazi:2019vch} for a fully controllable computation of horizon punctures from instantons in lower dimensional gravity). As we recently discussed, the correspondence between the information storage and  topological winding number may be related to a BH topological quantum memory protection from decoherence and noise; similarly to topological quantum computers \cite{Addazi:2020mnm}
  (see also Refs.\cite{Addazi:2020axm,Addazi:2020wnc} for its deep connection with {\it Holographic Naturalness} paradigm).
 This also means that for a large BH, the tunneling probability to a WH is expected to be 
  exponentially suppressed by its large entropy content, i.e. $e^{-N}$. 
 Interestingly, gravitational instantons can generate exponentially suppressed
 new correlators for SM particles
as an effect of fields' propagation on a non-trivial background \cite{I1,I2,I3,I4,I5}. 

Let us remark that GR admits BH solutions with topological baryon hairs as BH skyrme solutions \cite{BHS,BHS1,BHS2,BHS3}. 
Thus there are also {\it B-hairy instantonic configurations} as well because of gravitational instanton/BH correspondence. We never found such a remark on BH skyrme 
and gravitational instantons in any references in litterature but it appears to us a direct consequence of the gravitational instanton/BH correspondence.  
In BH Baryon/Skyrme solution, Baryon hair is preserved as a topological charge, associated to a topological conservation current, from quantum gravity effects
\cite{Dvali:2016sac}. This would suggest the emergence of topological selection rules which limit the possibility of B-violations 
from skyrme BHs: for example if the skyrme BH had $B=-1$ charge, 
a transition to a WH $(B=+1)$ would violate the B-number of $|\Delta B|=2$
and its topological current.


\begin{figure}
\begin{centering}
\includegraphics[scale=0.5]{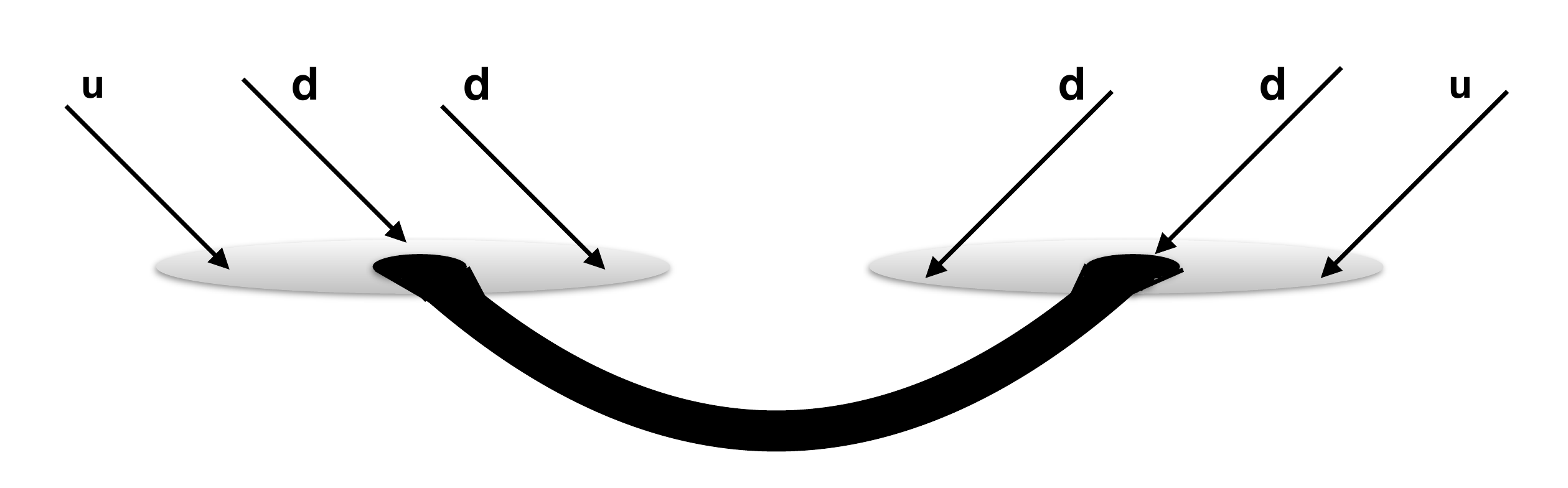}
\par\end{centering}
\caption{Gravitational skyrme instanton (black bridge) mediating a neutron-antineutron transition. (Anti)quarks composing the (anti)neutron enter in the (anti)skyrme region (gray) collectively coupled with 
the (WH)BH (anti)skyrme configuration.   }

\label{fig:jobInformationDialog}
\end{figure}


Nevertheless, following this logic, concerning neutrons,
one may consider a transition process as 
 $n + \widetilde{BH} \rightarrow \bar{n} +\widetilde{WH}$, 
where $\widetilde{BH},\widetilde{WH}$ denote $B=\mp 1$ hairy BH(WH) skyrme/antiskyrme (Fig.\ref{fig:jobInformationDialog}). 
Such a process is completely Baryon-number preserving.
We can think the neutron-$\widetilde{BH}$ as a $B=0$ bound state 
which can transform to a antineutron-$\widetilde{WH}$
system, without any violations of the topological $B$-number conservation. 
Let us also remark that such instantons do not destabilize protons and neutrons 
as $p\rightarrow l^{+}\pi^{0}$, $n\rightarrow l^{+}\pi^{-}$
and in general into any lepton/meson decay channel.
For example $n\,  \widetilde{BH}\rightarrow \widetilde{WH}\, l^{+}\, \pi^{-}$ does not 
preserve the baryon number, violating the B-topological protection
while the $n \, \widetilde{BH}\rightarrow \widetilde{BH}\, l^{+}\, \pi^{-}$ 
has the same transition probability of the SM-transition for energies well below the Planck scale.
Indeed, no any skyrme BH can have a lepton number since
originated from quarks, while lepto-quarks mediating proton decay would 
be problematic {\it di per se}, as well as beyond our no-new-particle approach.

\vspace{1cm}

\begin{figure}
\begin{centering}
\includegraphics[scale=0.5]{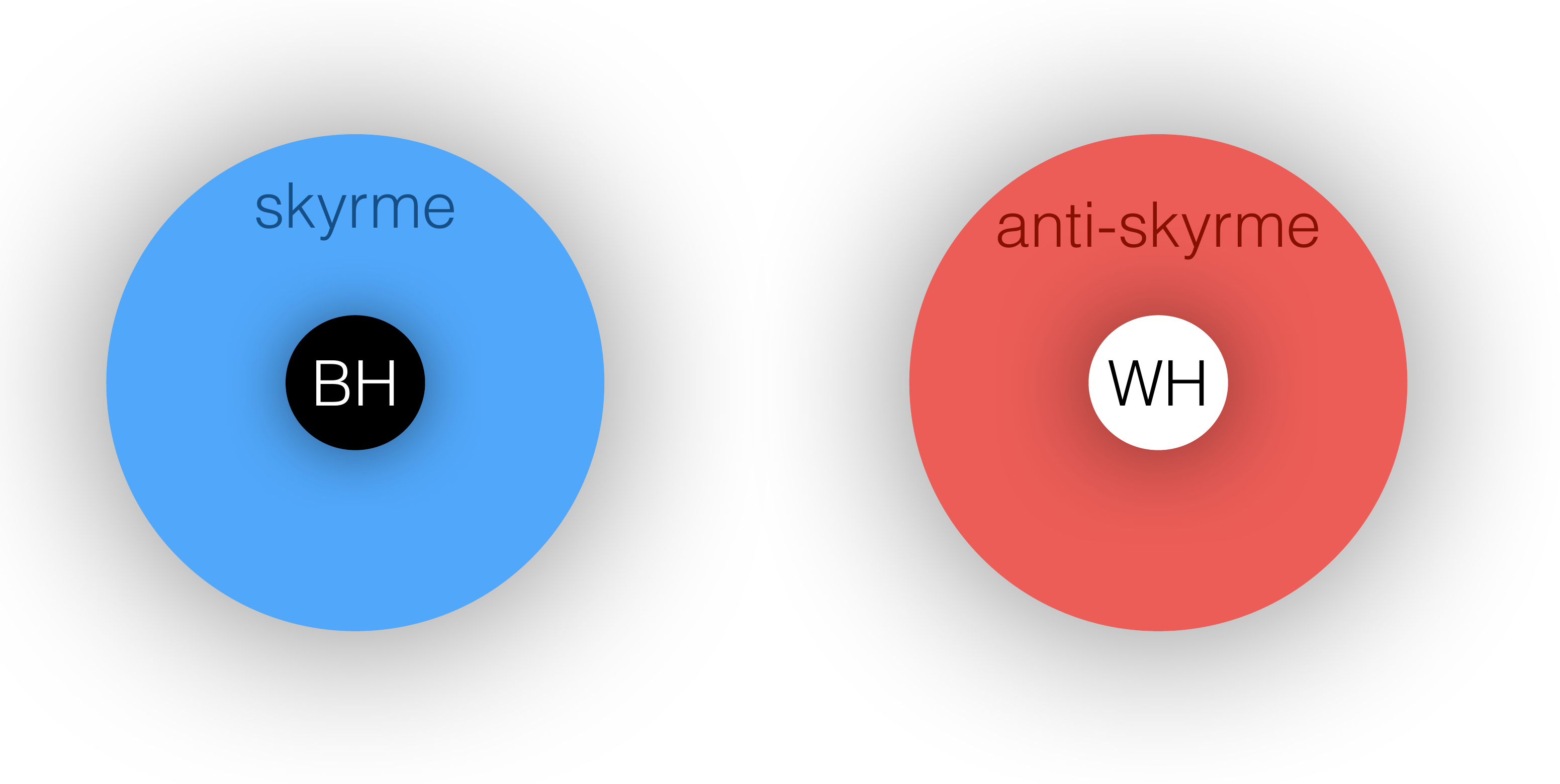}
\par\end{centering}
\caption{BH skyrme is a bound state of a BH (Black circle) and a Baryon charged soliton localized on a higher radius than the BH horizon (Blue region). The anti-configuration is a WH anti-skyrme bound state (WH as a White circle and anti-soliton in Red).  }

\label{fig:jobInformationDialog}
\end{figure}

\vspace{2cm}

Indeed, skyrme BHs can have topological B-hairs 
that are not localized in the the BH radius 
but on a larger skyrmion radius.
In other words, it is possible that micro-BH, even Planckian size ones,
can have Baryon/skyrme hairs spread on a much larger radius
as $R_{skyrme}>>R_{BH}$. 
In this case, the skyrme BH is coupled to the neutron 
rather than the single quarks if $R_{skyrme}\geq \Lambda_{QCD}^{-1}$ (see Fig.2).


The neutron-antineutron transition rate induced by quantum gravity effect 
maximally scales as the Planck scale. Nevertheless an entropic factor 
suppression associated to the BH-WH transition has also to be included:
\begin{equation}
\label{corr}
\langle n ,\, BH| \bar{n}\, , WH \rangle\sim e^{-\bar{I}_E}\sim e^{-N}\rightarrow \Gamma(n\rightarrow \bar{n}) \simeq M_{Pl}e^{-N}\,.
\end{equation}
where $N$ is the number of qubit stored in the micro-BH. 
Another way to see the $n\rightarrow \bar{n}$ 
is as an oscillation on a non-trivial vacuum state of quantum gravity.
The $n-\bar{n}$ transition can be induced 
by the gravitational skyrme instanton background 
as $\langle nn+h.c\rangle_{\mathcal{M}}$
$\rightarrow M_{Pl}e^{-I_{E}}nn+h.c$
where $\mathcal{M}$ is the non-trivial euclidean topology of the gravitational instantons.
In our case, the gravitational instanton is topologically equivalent
to a $S_{2}\times S_{2}$ double 2-sphere with two conical string defects,
-- since skyrme BH has conical angle defects in its metric as in the case of topological 
cosmic strings \cite{BHS}.

As it is well known, standard skyrme solutions, 
as well as string-like configurations eventually coupled to it, 
have fermionic zero modes.
The apparence of Majorana fermion modes 
from skyrme-like or string-like topological defects is almost ubiquitous in condensed matter and topological materials. 
In our case, Majorana fermionic zero modes 
appear out as modes which 
get a (Majorana) mass gap outside the skyrme radius 
while inside they are (Majorana) massless.
Thus we can consider a baryon preserving coupling 
of zero modes associated to the skyrme BH
and neutrons as 
\begin{equation}
\label{MI}
\mathcal{L}_i= \mu_{\tau n} \tau n+h.c. 
\end{equation}
where $\tau$ has inverse baryon number than the neutron
while $\mu_{\tau n}$  is a mass mixing. 
It is natural to assume as a missing singlet argument. We mean that no any symmetry principle seems to protect such a mass scale to be arbitrarily large.
that such a mass is only cut off by quantum gravity energy scale, 
i.e. we will assume that it is of the order of the Planck scale. 
Let us consider a pair of identical $\tau$ zero modes localized on the BH skyrme configuration. 
Integrating on zero mode space, the instanton transition 
amplitude has an extra contribution from  grassmanian integration
\begin{equation}
\label{S1}
\int d^2 \tau e^{-\mu_{\tau n} \tau n}\propto  nn\int d^2 \tau \tau\tau \rightarrow \frac{1}{M_{Pl}}\mu_{\tau n}^2 nn \,,
\end{equation}
where $\mu_{\tau n}$ is naturally thought as $\simeq M_{Pl}$, $d^{2}\tau$  because of two-zero modes 
and the $M_{Pl}$ in denominator comes from the path integral normalization as quantum gravity scale. In principle we can consider 
a larger zero mode space on the BH skyrme configuration but integral of higher species will only generate
$M_n (nn)(nn)^{N}/\Lambda^{3N}$ ($N\leq 1$) operators.

Incidentally, let us remark that such a mechanism exhibits an intriguing similarity 
with {\it exotic stringy instantons}, which also non-perturbatively generate new effective operators, integrating out supersymmetric modulini fields
(see Refs.\cite{Addazi:2014ila,Addazi:2015rwa,Addazi:2015goa} for specific discussions of neutron-antineutron case).

Eq.\ref{S1} is exponential suppressed by the gravitational skyrme instanton euclidean action
and thus we obtain a Majorana mass term for the neutron
as 
\begin{equation}
\label{S2}
M_{Pl} e^{-\bar{I}_{E}} nn +h.c\rightarrow M_{Pl} e^{-N} nn +h.c.
\end{equation}
In principle, we can also promote the B-number conservation
to a $U_{B}(1)$ global symmetry. 
Eqs.(\ref{S1},\ref{S2}) are subtly compatible with $U_{B}(1)$ symmetry: baryon transformation phase of $nn$ part, 
$nn\rightarrow e^{2i \alpha}nn$ is exactly compensated by the imaginary part acquired by the $e^{-\bar{I}_{E}}$
part as an effect of euclidean action part of fermionic zero modes. 
On the other hand the $U_{B}(1)$ is spontaneously broken by the presence of the BH skyrme 
configuration. 

In terms of quarks,
the effective coupling among the skyrme zero modes and quarks are
\begin{equation}
\label{effff}
\frac{\mu_{\tau n}}{\Lambda^3}\tau (udd)+h.c.
\end{equation}
where the effective $\Lambda$ cutoff will coincide with the QCD scale if the 
 the gravitational instanton is paired with an effective skyrme with radius around the QCD length, i.e. $\Lambda=\Lambda_{QCD}$.
 $\mu_{\tau n}$ is the same mass scale defined in Eq.\ref{MI}.  Integrating on the fermionic moduli space
 we obtain 
 \begin{equation}
 \label{kdd}
 \frac{\mu_{\tau n}^2}{M_{Pl}}e^{-\bar{I}_{E}}\frac{1}{\Lambda_{QCD}^{6}}(udd)^2+h.c.
 \end{equation} 
From this operator, we directly obtain Eq.\ref{S2}.

The emergent Majorana neutron mass term introduces 
an exponentially suppressed correlation between 
neutron and antineutron wave functions.
In terms of the neutron-antineutron effective Hamiltonian matrix 
we have 
\begin{equation}
\mathcal{H}=\begin{bmatrix} 
	E_{n} & M_{Pl} e^{-\bar{I}_{E}} \\
	M_{Pl} e^{-\bar{I}_{E}} & E_{\bar{n}} \\
	\end{bmatrix}
	\end{equation}
where in non-relativistic regime 
\begin{equation}
\label{aa}
E_{n}=m_{n}+\frac{p^{2}}{2m_{n}}+V_{n}\, , \,\,\,E_{\bar{n}}=m_{\bar{n}}+\frac{p^{2}}{2m_{\bar{n}}}+V_{\bar{n}}\, . 
\end{equation}

CPT invariance guarantees that $m_{n}=m_{\bar{n}}$, which in principle may be questionable if quantum gravity induces measurable quantum decoherence 
violating unitarity. Here we will assume that both CPT and Lorentz invariance are preserved by quantum gravity effects. 
In general, it is also assumed that neutron and antineutron respect the strong or weak equivalence principle
and that they feel the same gravitational field. 
In this case, if neutrons are not confined in a nucleus but just propagating on an external magnetic field,
we will just have $V_{n}=-V_{\bar{n}}=\mu \cdot B$
where $\mu$ is the magnetic moment of the neutron. 

Thus, for neutron-antineutron experiments 
the effective Hamiltonian would reduce to 
\begin{equation}
\mathcal{H}=\begin{bmatrix} 
	m_{n}+\mu \cdot B & M_{Pl} e^{-\bar{I}_{E}}  \\
	M_{Pl} e^{-\bar{I}_{E}}& m_{n}-\mu \cdot B \\
	\end{bmatrix}\, .
	\end{equation}





 
 Now, in order to have a testable neutron Majorana mass from gravitational instantons 
we should have 
\begin{equation}
M_{Pl} e^{-S} \sim 10^{-23}\, eV\, \rightarrow e^{-S} \sim 10^{-50}\, .
\end{equation}
Such a severe suppression corresponds to a reasonable entropy amount for a micro-BH:
$S \sim (\bar{r}/L_{Pl})^{2}\sim 115$ or so,
related to the same amount of planckian qubits. 
As it is known, gravitational instantons are interpreted as the appearence of virtual BH(WH) pairs 
and quantum mechanically also virtual pairs with skyrme hairs have to appear out. 
In our considerations we are forced to assume that 
the skyrme BHs contributing to the neutron Majorana mass 
cannot have a smaller amount of qubits than around $115$, 
otherwise the exponential suppression factor 
would not be competitive with the Planck mass scale 
and a too large Majorana mass would be generated out
from our mechanism. 
Nevertheless it is also natural to assume that a virtual BH
with baryon/skyrme hair necessarily has to contain a quibit number 
well larger than one since the very same skyrme solutions 
are generated out from a certain amount of degrees of freedom.


In this picture, the neutron-antineutron mixing 
seems to be related to a quantum entanglement of neutron and antineutron wave functions, 
which can be spatially separated by very large distances, 
rather than to a classical information exchange 
mediated by a new interaction.
This  is generated by the virtual BH pair, or a virtual BH-WH wormhole, 
as a new peculiar EPR=ER phenomenon \cite{Maldacena:2013xja}. 
In terms of kets, we can consider an entangled global state of neutron and antineutron 
for a certain position {\bf I} and position {\bf II} as 
$N(|n_I, \bar{n}_{II}\rangle+\epsilon e^{i\delta} |\bar{n}_{I},n_{II}\rangle)$, 
where $N=1/\sqrt{1+\epsilon^2}$. 
If {\it Alice} hypothetically measures the particle in {\bf I} and finds that it is a neutron,
then {\it Bob} will detect an antineutron in {\bf II}; and viceversa. 
In this prospective, in a neutron-antineutron experiment, 
if we observed a final antineutron state this means that we would actually observe 
an antineutron {\it quantum tele-transported} to our experimental detector localized on the Earth, eventually from 
very long distances. 
The probability of it is very tiny and it is 
$\langle n_{{\bf I}},\bar{n}_{{\bf II}}|\bar{n}_{{\bf I}},n_{{\bf II}}\rangle=N(\epsilon) \epsilon\sim e^{-N}$.
Such a quantum correlation is strongly suppressed if neutrons are bounded inside
nuclei: nuclear binding energy would create an effective mass gap splitting 
among neutrons and antineutrons completely suppressing 
quantum entanglements as a decoherence effect. 
Large external magnetic fields as well.





Now, let us move on another interesting possibility 
concerning Mirror matter. 
Let us consider a Mirror standard model (MSM)
with exactly the same gauge group of Ordinary SM
but eventually opposite parity \cite{M1,M2}.
In this way parity violations in the ordinary weak sector 
is secretly restored in M- sector. 
As it is known, if Mirror symmetry is preserved 
(not explicitly or spontaneously broken), 
a fast O- and M- neutron transition will be possible without any matter destabilization, 
i.e. $n-\bar{n}'$ oscillations. 
Let us define a separate O-
M- Baryon number for the O- and M- sectors
and that $n(\bar{n})$ carries $B=1(-1)$ while $n'(\bar{n}')$ carries a $B'=1(-1)$. 
Then we can consider new BH skyrme solutions where the topological conserved charge 
is a combination of $B,B'$. For simplicity let us consider a $B+B'$. 
Indeed, BH skyrme to a WH mirror skyrme can be envisaged 
as $n \, \widetilde{BH}\rightarrow \bar{n}'\widetilde{WH}$ 
preserving the $B+B'=0$ number, 
where $\widetilde{BH}\equiv \widetilde{BH}_{B+B'=B'=-1}$
and $\widetilde{WH}\equiv \widetilde{WH}_{B+B'=B=+1}$. 

Similarly to the neutron-antineutron case, this generates a 
mixing mass as 
\begin{equation}
\label{MI}
M_{Pl}e^{-\bar{S}'} \, n\, n'+h.c.\, 
\end{equation}
where $M_{Pl} e^{-\bar{S}'} $ has exclusion limits which are at lower energy scale
than the neutron Majorana mass of  
around 8th digits; but in terms of the exponential hierarchy with instanton entropy content 
is also around $100$ qu-bit planck digits. 
Such a BH skyrme can be thought as coupled to a {\it mirror Alice string}
which converts ordinary to mirror neutral particles,
eventually interpretable as a topological string defect from $SO(20)$ unification 
between O- and M- standard model sector \cite{Alice1,Alice2}. 
Similarly to the case of neutron Majorana mass, 
an operator in Eq.\ref{MI} can be generated by the 
mixing of fermionic zero modes related to the 
$B+B'$-hairy BH solution:
\begin{equation}
\mu_{\tau n}\tau n+ \mu_{\tau'n'}\tau' n'+h.c\, , 
\end{equation}
\begin{equation}
\int d\tau d\tau'e^{\mu_{\tau n}\tau n+ \mu_{\tau'n'}\tau' n'} \rightarrow \frac{1}{M_{Pl}}\mu_{\tau n}\mu_{\tau'n'}nn'+h.c.
\end{equation}
where $\tau,\tau'$ have O- and M- baryon numbers balancing O- and M- (anti)neutron
and we assume one and one localized on the BH skyrme configuration. 


As for neutron-antineutron, also $n-n'$ can be viewed as a quantum tele-trasportation phenomenon. 
Such a mechanism avoids any early Universe bounds on $n-n'$ from re-thermalization proposed 
in Ref.\cite{Babu:2021mjg} since not-related to any strong three O- M- quark collisions in early thermal bath.








\section{Baryon skyrme BHs and gravitational instantons}

The skyrme is a solitonic solution derived from a set of 
N quark flavors (or in the easiest case from two flavors)
having, in flat space-time, a baryon number as topological charge \cite{Skyrme,Skyrme1}: 
\begin{equation}
\label{kkk}
B=\int d^{3}x J_{0}\, , 
\end{equation}
where 
\begin{equation}
\label{jaj}
J_{\mu}=-\frac{1}{24\pi^{2}}\epsilon_{\mu\nu\rho\sigma} {\rm Tr} \big(U\partial^{\nu} U U^{-1}\partial^{\rho} U U^{-1}\partial^{\sigma} U)\, , 
\end{equation}
is the skyrme topological Chern-Simons current.
Above we defined $U$ as a $SU(N)$ flavor matrix 
that can be written as 
\begin{equation}
\label{UU}
U=e^{i\Pi_{a}(x)\tau_{a}/F}
\end{equation}
where $F$ is a characteristic constant, 
$\tau_{a}$ group matrices of $SU(N)$ 
and $\Pi_{a}$ are meson fields related to $SU(N)$. 
In case of two quarks, $SU(N)$ is just $SU(2)$,
$\tau_{a}$ are the Pauli matrices
$\Pi_{a}$ correspond to pion fields, $F$ to the pion decay constant. 
In particular, the Skyrme effective lagrangian can be written in terms of $U$
as in Ref.\cite{Skyrme} and the skyrme current in Eq.\ref{jaj} can be easily derived from it. 
Typically, in the $SU(2)$ flavor case, syrmion has 
a typical size and mass controlled by the pion decay constant and the electric charge 
as $L=e^{-1}F^{-1}$ and $M=Fe^{-1}$. 

The action originating a BH skyrme solution reads as 
\begin{equation}
\label{lla}
I=I_{m}+\int_{\Sigma} \sqrt{-g}d^{4}x\Big(\frac{1}{16\pi}(R-2\Lambda) \Big)
\end{equation}
$$+\frac{1}{8\pi}\int_{\partial \Sigma}\sqrt{h}d^{3}x(K-K^{0})\, . $$
where $I_{m}$ includes 
\begin{equation}
\label{matter}
\int d^{4}x\sqrt{-g}{\rm Tr}\Big(\frac{F^{2}}{16}X^{\mu}X_{\mu}+\frac{1}{32e^{2}}Y^{\mu\nu}Y^{\mu\nu}\Big) \, , 
\end{equation}
\begin{equation}
\label{kakll}
X_{\mu}=U^{-1}\nabla_{\mu} U, \,\,\,Y_{\mu\nu} =[X_{\mu},X_{\nu}]\, .
\end{equation}

In the case of a skyrme BH, 
we can consider a two sphere with radius $r$ closing enclosing 
the BH horizon as $r>>R_{BH}$ ($R_{BH}$ the BH event horizon):
\begin{equation}
\label{BB}
B=\int_{S_{2}}dx^{\mu} \wedge dx^{\nu}\mathcal{J}_{\mu\nu}
\end{equation}
with $\mathcal{J}_{\mu\nu}$ the topological current coupled to 
the embedding coordinates 
$x$. Indeed $J$ in Eq.\ref{jaj} represents the Hodge-dual of 
the exterior derivative of the two-form 
$\mathcal{J}$. 

As remarked in Ref.\cite{Dvali:2016sac}, one can show with a simple argument 
that the baryon topological hairs can be detected through 
a memory phase shift effect {\it a la} Aharonov-Bohm. 
For instance, one can consider 
a path string loop surrounding a skyrmion BH solution
considering an effective coupling $g$ between the string and the skyrmion 
$g \int dX^{\mu} \wedge dX^{\nu} \, \mathcal{J}_{\mu\nu}$.
Indeed, this can be the case of string flux-tubes of gauge fields, 
cosmic strings of fundamental strings. 
In this case, any physical processes 
in which the string world-volume encloses the syrmion BH
lead to the Aharonov-Bohm phase shift $\Delta \Phi=2\pi g$. 

Let us also remark that in a generic curved space-time, B has the integral form as 
\begin{equation}
B=\int_{\Sigma} d^{3}x\sqrt{-g}J^{0}=\int_{\partial \Sigma}dS J^{a}n_{a}\, 
\end{equation}
where $dS$ is the surface element of $\partial \Sigma$ which delimits the volume $\Sigma$
while $n^{a}$ is the unit normal vector equipping the surface. 
Considerations about Aharonov-Bohm shifts induced by B-hair, introduced above, can be generalized in curved space-time.

As found by many authors, spherically symmetric skyrme BH solutions 
can be found from a skyrme lagrangian coupled to Einstein-Hilbert equation \cite{BHS,BHS1,BHS2,BHS3}. 
A large class of skyrme BHs are basically a bound state of a skyrme soliton and a ordinary 
spherically symmetric BH. In particular the ADM mass of the sperically symmetric skyrme BH 
is in many cases just the sum of the skyrme solitonic energy, the mass associated to the BH horizon
and their biding energy:
\begin{equation}
M_{ADM}=M_{s}+M_{h}+E_{B}\, 
\end{equation}
where $M_{s}$ is the mass associated to the skyrme configuration, $M_{h}$ is the one 
contained within the BH horizon and $E_{B}$ is a binding energy of the two. 
The three depends on the specific metric profiles -- see for example Ref.\cite{NI}. 
In our case, we consider $M_{h}\simeq 10 M_{Pl}>> M_{s}$, while $R_{h}\simeq 10 L_{Pl}<<R_{s}\simeq L_{QCD}$
with $R_{h,s}$ BH horizon and soliton radii respectively, $L_{QCD}=1/\Lambda_{QCD}$. 
Such a case is naturally possible since the solitonic configuration has an energy density much smaller than 
the BH. Indeed this also implies that most of the entropy content of the bound state is (holographically)
localized on the BH horizon, i.e. $S_{BH}\sim (R_{h}/L_{PL})^{2}>> S_{s}$ where $S_{s}$ is the entropy content of the solitonic configuration.

Let us consider the partition function of the BH/skyrme solution:
\begin{equation}
\label{jajaakl}
Z=\int \mathcal{D}g \mathcal{D} \Psi\, e^{-I_{E}}
\end{equation}
where $I_E$ is the eucliden action of the system of Eq.\ref{lla}, $g$ is the metric tensor and $\Psi$ in general indicates all matter/gauge fields. 
As we know, gravitational instantons are saddle classical solutions of the Euclidean action. 
Gravitational instanton can be classified from Euler numbers and signature
\begin{equation}
\label{jkkaaaa}
\chi=\nu+B_{2}^{+}+B_{2}^{-}=\frac{1}{128\pi^{2}}\int d^{4}x\sqrt{g}R_{abcd}R_{a'b'c'd'}\epsilon^{aba'b'} \epsilon^{cdc'd'}\, , 
\end{equation}
\begin{equation}
\label{taub}
\tau=B_{2}^{+}-B_{2}^{-}=\frac{1}{96\pi^{2}}\int d^{4}x\sqrt{g}R_{abcd}R_{a'b'}^{cd} \epsilon^{aba'b'}\, , 
\end{equation}
where $\nu=2$ for compact manifolds and $\nu=1$ for non-compact ones
and 
$B_{2}^{\pm}$ are the second Betti numbers 
of the harmonic and anharmonic two forms, respectively. 
In case of gravity coupled to matter, without any topological defects in matter sector, 
one can individuate three main topological building blocks 
as 
$S_{2}\times S_{2}$, $CP^{2}$ and $K^{3}$;
but 
$S^{2}\times S^{2}$ with $(\chi,\tau)=(4,0)$
is the dominant class \cite{I1,I2,I3,I4,I5}. 
As mentioned, $S^{2}\times S^{2}$ is interpreted as related to 
virtual BH pairs. 
In case of skyrme configuration, 
topology of  $S^{2}\times S^{2}$ is eventually 
punctured by opening string-like angles.
Nevertheless this topological difference does not lead to 
any substantial difference or suppression/enanchement 
compared to the $S^{2}\times S^{2}$ configuration. 
As it is well known, the entropy related to the BH configuration is related 
to the partition function and the BH skyrme case follows the same thermodynamical rule:
\begin{equation}
\label{kkakaoo}
S={\rm log}Z+\beta E\, \simeq S_{h}\, 
\end{equation}
which, as mentioned before, is dominated by the entropy stored on the horizon surface. 
The $\widetilde{BH}-\widetilde{WH}$ transition probability is 
exponentially suppressed as the action evaluated on the gravitational skyrme instanton
as 
\begin{equation}
\label{Gamma}
\Gamma=Ae^{-I_{\mathcal{M}}}\sim e^{-c\, (R_{h}/L_{Pl})^{2}}
\end{equation}
where $A$ also includes quantum loop corrections, 
$I_{\mathcal{M}}$ is the action of the gravitational instanton
with topology $\mathcal{M}\equiv S_{2}\times S_{2}-\{p\}$ with $\{p\}$ puncture set
$c$ a numerical coefficient not relevant for our discussions. Similar considerations for the 
more standard $S_2\times S_2$ case can be found in Refs.\cite{II1,II2,II3,II4}.
The rate can also be expressed in the terms of the imaginary part of the partition function 
as 
\begin{equation}
\label{kak}
\Gamma=-\frac{1}{\sqrt{3}\pi}\frac{1}{R_{h}}\frac{{\rm Im}Z[\mathcal{M}]}{Z[S_{4}]}\, . 
\end{equation}
where $S_{4}$ is the Euclidean four-dimensional sphere. 

\vspace{0.5cm}

\section{Conclusions and remarks.} 

In this paper, we considered a novel mechanism 
generating a Majorana mass for the neutron 
from virtual skyrme/baryon black holes. 
In particular,  quantum gravity effects
may subtly survive at the Fermi scale generating 
a Majorana mass gap for the neutron rather than 
having a direct effect on single quark constituents.
Indeed, GR admits
BH/skyrme bound systems which 
have the following properties;
i) they have a topological Baryon hair 
which 
generates a Aharanov-Bohm like effect
on SM baryons;
ii) the BH and skyrme radii can be very different
and the skyrme size can be much larger than the BH diameter;
iii) Baryon violating transitions generated by such configurations 
follow selection rules which allow for proton stability.
We also showed that neutron mirror neutron mixing can be generated
from virtual BHs with O- and M- skyrme/baryon hairs. 

Contrary to typical mechanisms to generate a $n-\bar{n}$ or a $n-n'$ transition, 
here these mixings are effects of long-range correlations
coherently surviving from Planck world to larger length scales. 
A neutron Majorana mass of about $10^{-23}\, eV$ is related to a long-range correlation 
between neutron and antineutron. This seems to suggest that such gravitational instantons, 
related to wormholes, can have a long-extension of warp throats, even if the corresponding 
BH(WH) sizes are small as few times the Planck length. 
Contrary to standard interactions, such a correlation is not suppressed by the distance but by the semiclassical probability 
of BH-WH skyrme/anti-skyrme tunneling, in turned exponentially suppressed as the BH(WH) entropy. 
This can be effectively viewed as a long-range gravitational entanglement between neutron and antineutron 
which may be interpreted as an intriguing manifestation of $ER=EPR$ conjecture \cite{Maldacena:2013xja}.
Indeed, this gravitational entanglement will be highly (but not fully) suppressed by decoherence effects if neutrons are confined in nuclei
or in presence of strong external magnetic fields. 
Similarly, if mirror neutrons existed, a $n-n'$ long-range entanglement mediated by O/M skyrme BHs could be also envisaged.
In this sense, quantum gravity has unexpected topological portals from matter to anti-matter (and viceversa) through {\it apparent} 
$|\Delta B|=2$ and $|\Delta B|=1$ violating transitions -- another strong motivation towards neutron-antineutron searches.

\vspace{0.6cm}

\noindent

\vspace{0.5cm}

{\bf Acknowledgements}
Our work is supported by the Talent Scientific Research Program of College of Physics, Sichuan University, Grant No.1082204112427
\& the Fostering Program in Disciplines Possessing Novel Features for Natural Science of Sichuan University,  Grant No. 2020SCUNL209
\& 1000 Talent program of Sichuan province 2021.

\end{document}